# Carbon tips as electrodes for single-molecule junctions

Andres Castellanos-Gomez[1,*,†], Stefan Bilan[2], Linda A. Zotti[2], Carlos R. Arroyo[1,†], Nicolás Agraït[1,3,4], Juan Carlos Cuevas[2,*], Gabino Rubio-Bollinger[1,3,*]

[1] Departamento de Física de la Materia Condensada (C–III). Universidad Autónoma de Madrid, Campus de Cantoblanco, 28049 Madrid, Spain.

[2] Departamento de Física Teórica de la Materia Condensada (C–III). Universidad Autónoma de Madrid, Campus de Cantoblanco, 28049 Madrid, Spain.

[3] Instituto Universitario de Ciencia de Materiales "Nicolás Cabrera". Campus de Cantoblanco, 28049 Madrid, Spain.

[4] Instituto Madrileño de Estudios Avanzados en Nanociencia IMDEA-Nanociencia, 28049 Madrid, Spain.

We study electron transport through single-molecule junctions formed by an octanethiol molecule bonded with the thiol anchoring group to a gold electrode and the opposing methyl endgroup to a carbon tip. Using the scanning tunneling microscope based break junction technique, we measure the electrical conductance of such molecular junctions. We observe the presence of well-defined conductance plateaus during the stretching of the molecular bridge, which is the signature of the formation of a molecular junction.

Understanding electron transport through a single molecule is the main goal in molecular electronics. A primary aim is to find ways to form a reliable mechanical and electrical connection between the molecule under study and the macroscopic electrodes. The mechanical and electrical properties of a molecular junction can be tuned by a clever design of the molecular structure, but also by a proper selection of the material of the electrodes. So far, gold has been the main material of choice for the electrodes in Scanning Tunneling Microscope break junction (STM-break junction) experiments because gold is chemically inert, even in air, while molecules can be bound to it with well characterized linker groups such as thiols [1-3] or amines [1-3]. However, the bond formation between gold electrodes and methyl terminated molecules such as alkane(mono)thiols, although possible in certain electrochemical environments [4], has proven to be elusive in break-junction experiments[5,6]. This explains why the use of gold as electrodes in single-molecule electronics has limited the variety of molecular systems that can be studied with STM-break junction experiments. In this sense, it would be highly desirable to widen the list of materials that can be used as electrodes [2,7]. Carbon is obviously a good candidate as an electrode material in molecular electronics because it offers the possibility to contact a variety of molecules without the need of anchoring groups, it is also relatively inert, and it forms structures like graphene [8,9] and nanotubes [10] that can be used as nanoelectrodes.

---

* E-mail: a.castellanosgomez@tudelft.nl , juancarlos.cuevas@uam.es , gabino.rubio@uam.es

† Present address: Kavli Institute of Nanoscience, Delft University of Technology, Lorentzweg 1, 2628 CJ Delft, The Netherlands.

In this work, we have explored the suitability of carbon-based tips as electrodes to form molecular junctions. Using the STM-break junction technique we have measured the electrical conductance of several hundred octanethiol-based single-molecule bridges ($CH_3$-$C_7H_{14}$-SH) in which the thiol anchoring group is bound to a gold electrode and the methyl group is linked to a carbon electrode. Most of the conductance traces measured while stretching the molecular junction show a marked plateau which is the signature of the formation of molecular junctions [11,12].

In order to form single-molecule junctions with a carbon electrode we used a homebuilt scanning tunneling microscope (STM) [13] with a carbon fiber tip. These tips have been characterized for their use in STM proving that the carbon tip apex is not contaminated [14]. We have also ruled out the possibility that the carbon tip is contaminated or wetted by gold atoms of the surface [15].

Single-molecule junctions are obtained by repeatedly forming and breaking the contact between the tip and a gold substrate covered with a self-assembled monolayer (SAM) of molecules. We choose octanethiol molecules as a model system because it has the following advantages. First, SAMs of octanethiol molecules can be routinely obtained and are well-characterized by STM studies [16]. Second, octanethiol is the shortest, and thus most conductive, alkanethiol that forms highly ordered standing-up phase monolayers [17]. Third, the strong bonding between the thiol group and the gold substrate prevents the tip from pulling the molecules off the substrate during the formation and breaking of the contacts.

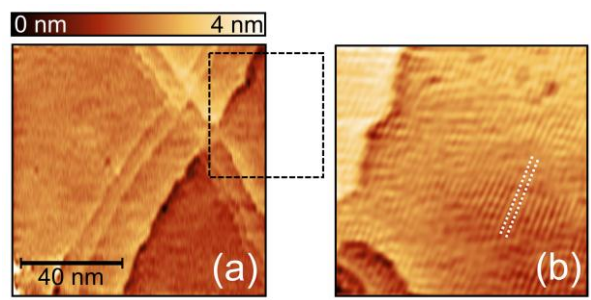

**Figure 1.** (a) Constant current STM topography image of atomically flat Au (111) terraces covered by a dense octanethiol monolayer ($I_{tunnel}$ = 1.6 nA, $V_{tip}$ = -300 mV). (b) Detail of the region marked by the dashed square in (a). The parallel dotted white lines indicate the direction of the characteristic stripe-like structure.

We characterized the SAM using STM topography images at constant current with a tunneling resistance of 187 MΩ. The images in Figure 1 reveal large areas of triangular-shaped atomically flat terraces separated by sharp monatomic steps, distinctive of Au (111) reconstructed surfaces. The higher detail image Figure 1(b) shows the characteristic stripe-like structures produced by the self-assembly of the alkanethiols into periodic molecular domains [18].

After the characterization of the SAM surface we formed single molecule contacts with the carbon tip. We start from a non-contact situation with a tunneling resistance higher than 200 MΩ at an applied fixed tip bias of 100 mV. In order to approach the tip to the sample in a controlled way the STM current feedback setpoint is lowered to a value of 50 nA, corresponding to a resistance of 2 MΩ, in a time lapse of 200 ms. The tip approach is sufficient to bring the carbon tip into gentle contact with the SAM but small enough not to touch the gold surface. After the shallow contact is established the current feedback is turned off and the tip is retracted 2 nm away from the surface, at a rate of 60 nm/s. During this retraction we measure the current at a fixed bias. A comparison between the STM topographic images before and after the indentation demonstrates that deeper indentation of the tip causes irreversible damage of the SAM. The indentation-retraction cycle is completed by turning the STM feedback loop on. This cycle is repeated 32 times at a predetermined spot of the surface. We have studied 20 different spots of the sample separated by approximately 50 nm.

Several traces of conductance as a function of tip retraction distance $G(z)$ are shown in Figure 2(a). The conductance traces show plateaus at specific values which are the signature of the formation of molecular junctions [17]. The plateaus are separated by a sharp conductance drops related to the rupture of the molecular bridge. The conductance plateaus do not always occur at the same conductance values because of variations in the microscopic details of the molecular arrangement between the electrodes. A statistical analysis over a large number of molecular junctions is thus necessary. To overcome junction-to-junction fluctuations we have performed a statistical analysis in which all junction realizations, without sifting conductance traces, are represented as a histogram [13]. Figure 2(b) shows the conductance histogram (dark blue) with a broad hump centered at $G = 8 \times 10^{-6}\ G_0$, where $G_0 = 77.5\ \mu S$ is the conductance quantum, and a smooth and monotonically decreasing background. The hump is associated with the presence of conductance plateaus in individual traces and the background with tunneling conduction. Indeed, the individual traces in Figure 2(a) show that, except at the plateau, the conductance decreases exponentially as a function of the tip separation although its decay constant differs before and after the last conductance plateau. The mean value of the apparent tunneling barrier height before the plateau is 0.6-1.0 eV, which can be attributed to tunneling through the SAM. After the molecular bridge breaks, the conduction mechanism is due to vacuum tunneling under ambient conditions [19] and the mean barrier height is 1.0-2.4 eV. The fact that the apparent tunneling barrier height is lower when the tip is in contact with the SAM is in good agreement with previous work in which a lowering of the apparent barrier height is suggested due to the elastic surface deformation induced by the repulsive forces during the tip-sample contact [2]. Therefore, the background in the conductance histogram can be attributed to vacuum tunneling under ambient conditions for conductance values below the peak (around $G < 3 \times 10^{-6}\ G_0$) and to tunneling through the SAM for conductance above the peak (around $G > 1.5 \times 10^{-5}\ G_0$).

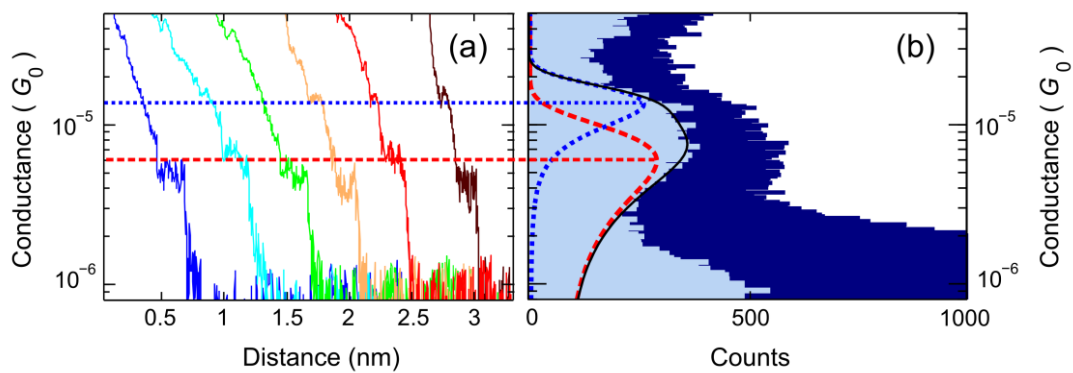

**Figure 2.** (a) Conductance traces (shifted horizontally by 0.5 nm for clarity) showing characteristic conductance plateaus. Below a conductance of $1.5 \times 10^{-6}\ G_0$ the measurements are limited by the noise level of the current amplifier. (b) Conductance histogram built from 740 traces (dark blue) and corrected histogram (light blue) after subtracting the tunnelling contribution, following the procedure described in ref. [20]. The corrected histogram shows two conductance peaks. The dashed red and the dotted blue curves are the two Gaussians and the solid black curve is their sum. The maximum of the conductance peaks mark the two most probable molecular configurations (horizontal dashed/dotted lines).

We followed the procedure described in ref. [21] to remove the background tunneling contribution from the histogram in order to better resolve the structure of the conductance hump. Since the slope of the conductance is lower on the plateaus than in the tunneling regions, the tunneling contribution to the conductance histogram can be removed by considering only intervals of the individual $G(z)$ trace whose exponential decay constant is lower than a threshold value. The corrected histogram in Figure 2(b) (light blue) is obtained in this way using a threshold value which corresponds to an apparent barrier height of 450 meV. The result is that the background due to the tunneling contribution has been reduced by a factor of 3-4 and the structure of the broad hump in conductance can be better resolved.

We find that the broad hump can be fitted to the sum of two Gaussian peaks in a linear conductance scale whose centers are located at $G_1 = (5.9 \pm 4.1) \times 10^{-6}\ G_0$ and $G_2 = (1.3 \pm 0.5) \times 10^{-5}\ G_0$. The presence of multiple peaks in the histogram in previous STM break junction experiments on molecular junctions has been attributed to a varying number of molecules contributing to the transport [21]. The fact that the value of $G_2$ is twice that of $G_1$ suggests that plateaus at conductance $G_2$ correspond to electron transport through two simultaneously connected molecules each of which has a conductance $G_1$.

We compare the conductance value of these gold-octanethiol-carbon single molecular junctions with other model systems. The conductance through octanethiols has been studied using gold electrodes in configurations such as 2D nanoparticle arrays [13] and nanopores [22] with a relatively large gold-octanethiols-gold junctions (400 – 1000 molecules are involved). The conductance per molecule has been found to be in the range of $(8-70) \times 10^{-8}\ G_0$ [23] which is more than an order of magnitude lower than the value we have obtained for gold-octanethiol-carbon junctions. On the other hand, break junction experiments with octanethiol molecules and gold electrodes show that not only the conductance is low, but so is the probability of the formation of molecular junction [22,23]. The usual approach to provide both good electrical contact and strong chemical bonding to gold electrodes is to functionalize the molecules with thiol anchoring groups on both ends. In single-molecule experiments with octanedithiols the reported conductance is in the range of $(1-25) \times 10^{-5}\ G_0$ [2,7,24]. The corresponding value for the conductance of the carbon tip contacted octanethiols is comparable in spite of the lack of a specific anchoring group. Thus the comparison with other model systems supports the suitability of carbon fiber tips to contact methyl terminated molecules.

The appearance of a plateau in the conductance traces suggests the formation of a chemical bond between the alkane chain and the carbon tip. However, it is not easy to say *a priori* how a methyl-terminated chain can bind to a pure carbon-hydrogen structure. For this reason and in order to shed some light on the formation mechanism of our molecular junctions, we have performed simulations within the framework of density functional theory (DFT) to find/propose a feasible scenario for the formation of a stable alkane-carbon contact [25].

In conclusion, we have used carbon fiber tips as electrodes in an STM-break junction configuration to form single-molecule junctions with octanethiol deposited on a gold surface. We find that carbon tips provide a proper mechanical linking to the methyl group allowing the routine formation of single-molecule bridges. The conductance traces during the stretching of the bridges show well defined plateaus that are reflected in the conductance histogram as a prominent peak at $5.9 \times 10^{-6}\ G_0$ which corresponds to the conductance of a single molecule. This value is comparable to that of octanedithiol between gold electrodes. It is important to emphasize that these carbon tips can be suitable candidates to contact a variety of organic molecules and they can also be combined with other substrate materials including carbon itself to form purely organic single-molecule devices.

**Acknowledgements**

The authors wish to acknowledge the help of R. H. M. Smit and E. Leary for carefully reading the manuscript and for useful discussions. A.C-G. acknowledges fellowship support from the Comunidad de Madrid (Spain). This work was supported by MICINN (Spain) through the programs MAT2008-01735 and CONSOLIDER-INGENIO-2010 CSD-2007-00010, by Comunidad de Madrid through the program NANOBIOMAGNET S2009/MAT1726, by the EU through the networks BIMORE (MRTN-CT-2006-035859) and by the EC through the network FP7 ITN "FUNMOLS" Project Number 212942.

**References**

[1] M. A. Reed, C. Zhou, C. J. Muller, T. P. Burgin, and J. M. Tour, Science **278** (5336), 252 (1997).
[2] X. Li, J. He, J. Hihath, B. Xu, S. M. Lindsay, and N. J. Tao, J. Am. Chem. Soc. **128** (6), 2135 (2006).
[3] C. Li, I. Pobelov, T. Wandlowski, A. Bagrets, A. Arnold, and F. Evers, J. Am. Chem. Soc. **130** (1), 318 (2008).
[4] E. Wierzbinski and K. Slowinski, Langmuir **22** (12), 5205 (2006).
[5] F. L. X. Chen, J. H. Z. Hihath, and N. Tao, J. Am. Chem. Soc. **128** (49), 15874 (2006).
[6] S. Y. Quek, L. Venkataraman, H. J. Choi, S. G. Louie, M. S. Hybertsen, and J. B. Neaton, Nano Lett. **7** (11), 3477 (2007).
[7] S. Wu, M. T. Gonzalez, R. Huber, S. Grunder, M. Mayor, C. Schonenberger, and M. Calame, Nature Nanotech. **3** (9), 569 (2008).
[8] M. Kiguchi, O. Tal, S. Wohlthat, F. Pauly, M. Krieger, D. Djukic, J. C. Cuevas, and J. M. van Ruitenbeek, Phys. Rev. Lett. **101** (4), 46801 (2008).
[9] R. H. M. Smit, Y. Noat, C. Untiedt, N. D. Lang, M. C. van Hemert, and J. M. van Ruitenbeek, Nature **419** (6910), 906 (2002).
[10] H. W. C. Postma, Nano Lett. **10** (2), 420 (2010).
[11] X. Guo, A. A. Gorodetsky, J. Hone, J. K. Barton, and C. Nuckolls, Nature Nanotech. **3** (3), 163 (2008).
[12] S. Roy, H. Vedala, A. D. Roy, D. Kim, M. Doud, K. Mathee, H. Shin, N. Shimamoto, V. Prasad, and W. Choi, Nano Lett. **8** (1), 26 (2008).
[13] B. Xu and N. J. Tao, Science **301** (5637), 1221 (2003).
[14] R. H. M. Smit, R. Grande, B. Lasanta, J. J. Riquelme, G. Rubio-Bollinger, and N. Agrait, Rev. Sci. Instrum. **78** (11), 113705 (2007).
[15] See supplementary material at [URL will be inserted by AIP] for a detailed description of the tip preparation and a control experiment to rule out the wetting of the tip apex by gold atoms of the surface.
[16] A. Castellanos-Gomez, N. Agrait, and G. Rubio-Bollinger, Nanotechnology **21** (14), 145702 (2010).
[17] M. Sharma, M. Komiyama, and J. R. Engstrom, Langmuir **24** (18), 9937 (2008).
[18] G. E. Poirier and E. D. Pylant, Science **272** (5265), 1145 (1996).
[19] M. T. Gonzalez, S. Wu, R. Huber, S. J. van der Molen, C. Schönenberger, and M. Calame, Nano Lett. **6** (10), 2238 (2006).
[20] Y. A. Hong, J. R. Hahn, and H. Kang, J. Chem. Phys. **108**, 4367 (1998).
[21] J. L. Xia, I. Diez-Perez, and N. J. Tao, Nano Lett. **8** (7), 1960 (2008).
[22] J. Liao, L. Bernard, M. Langer, C. Schönenberger, and M. Calame, Adv. Mater. **18** (18), 2444 (2006).
[23] W. Wang, T. Lee, and M. A. Reed, Phys. Rev. B **68** (3), 035416 (2003).
[24] W. Haiss, R.J. Nichols, H. van Zalinge, S.J. Higgins, D. Bethell, and D.J. Schiffrin, PCCP **6** (17), 4330 (2004).
[25] See supplementary material at [URL will be inserted by AIP] for a detailed description of the DFT simulations.

# Supplementary material:

## Carbon tips as electrodes for single-molecule junctions

Andres Castellanos-Gomez[1,\*,†,] Stefan Bilan[2], Linda A. Zotti[2], Carlos R. Arroyo[1,†], Nicolás Agraït[1,3,4],
Juan Carlos Cuevas[2,\*], Gabino Rubio-Bollinger[1,3,\*]

1 Departamento de Física de la Materia Condensada (C–III). Universidad Autónoma de Madrid, Campus de Cantoblanco, 28049 Madrid, Spain.
2 Departamento de Física Teórica de la Materia Condensada (C–III). Universidad Autónoma de Madrid, Campus de Cantoblanco, 28049 Madrid, Spain.
3 Instituto Universitario de Ciencia de Materiales "Nicolás Cabrera". Campus de Cantoblanco, 28049 Madrid, Spain.
4 Instituto Madrileño de Estudios Avanzados en Nanociencia IMDEA-Nanociencia, 28049 Madrid, Spain.
[*] E-mail: a.castellanosgomez@tudelft.nl , juancarlos.cuevas@uam.es , gabino.rubio@uam.es
[†] Present address: Kavli Institute of Nanoscience, Delft University of Technology, Lorentzweg 1, 2628 CJ Delft, The Netherlands.

**Carbon tip preparation**

Carbon fiber tips are prepared from freshly cut individual carbon fibers, obtained from a commercially available carbon fiber rope [1], and are mounted in a home-built scanning tunneling microscope [2] by gluing them to the tip holder, protruding 100 μm, with conductive epoxy[3]. The microscopic structure of the tip is composed by graphitic planes aligned parallel to the fiber longitudinal axis [3] yielding high electrical conductivity $\sigma = 7.7 \times 10^4$ S/m. Simultaneous measurements of the tunneling current and the force as a function of the separation between the tip and a gold surface indicate that the carbon tip apex is not contaminated [4]. This makes carbon tips good candidates for electrically contacting molecules in air or liquid environments.

**Sample preparation**

The molecules were deposited onto a gold substrate which was initially treated with piranha solution and then flame annealed to prepare a flat reconstructed Au (111) surface. The substrate was incubated for 12 hours in a 1 mM solution of octanethiol (Aldrich) in toluene, then rinsed and sonicated in pure toluene and subsequently dried in a stream of helium gas. These deposition conditions are well known to yield a densely packed self-assembled monolayer [5].

**Carbon–gold nanocontacts**

We have studied the formation of the contact between a carbon fiber tip and a clean gold (111) surface. We compare the measurements of the electrical conductance (*G*) trace during two different indentation experiments on a clean gold (111) surface: one in which we use a gold tip (hereafter gold-gold) and another one in which we use a carbon fiber tip (here after carbon-gold). Among other findings, such a comparison allows us to rule out wetting processes, by gold atoms of the surface, of the carbon fiber tip.

For the gold-gold indentation experiment we reproduce the well-known results reported in the literature [6], that is, that the conductance changes stepwise during the indentation, a behavior which is related to sudden atomic rearrangements taking place at the narrowest part of the contact. In particular, it is observed that the well-defined smallest contact, formed by a single gold atom, has a conductance value of $G_0 = 2e^2/h$. Another signature of gold nanocontacts is that the transition from electron tunneling to contact regime is abrupt [7], which is due to a sudden mechanical jump to contact process [8]. The conductance plateaus do not always occur at the same conductance values because of variations in the microscopic details of the atomic arrangement in the electrodes, making a statistical analysis over a large number of indentation cycles necessary. We have therefore analyzed thousands of indentation traces to build a two-dimensional conductance histogram [9] (Figure S1a), which displays a well-known stepped structure. In addition, the one-dimensional conductance histogram is shown in Figure 1b and can be considered a fingerprint of gold-gold nanocontacts, as its peaked structure varies from metal to metal [10].

---

[3] Silver loaded epoxy adhesive, purchased from RS with product number 186-3616.

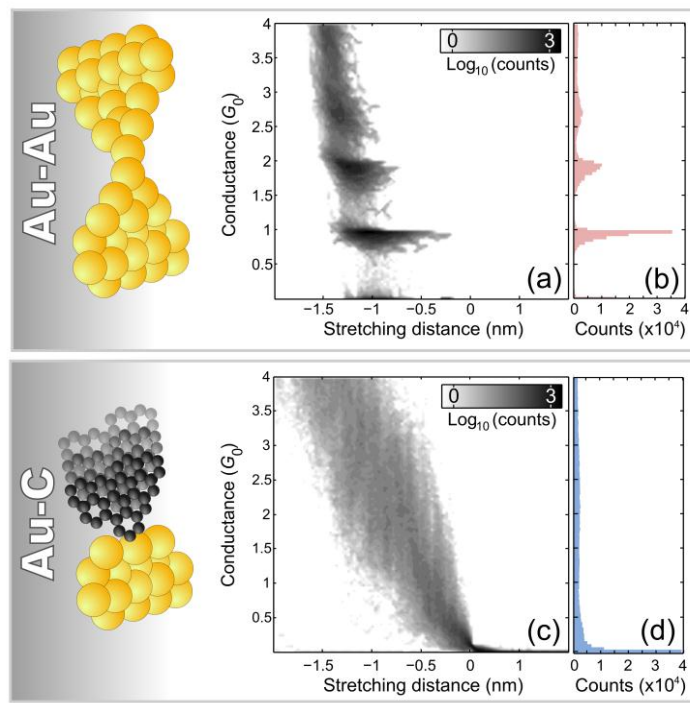

**Figure S1:** (a) and (c) are two-dimensional conductance histograms from 1000 consecutive breaking traces of gold-gold and carbon-gold nanocontacts, respectively. Areas with highest counts (darkest areas in this plot) represent the most typical breaking behaviour of the individual nanocontacts. (b) and (d) show the corresponding one-dimensional conductance histograms for the gold-gold and carbon-gold nanocontacts respectively.

For the carbon-gold indentation experiment the individual conductance traces neither show a stepped structure nor an abrupt jump at a well-defined conductance value. This results in a featureless two-dimensional conductance histogram (Figure S1c) as well as a one-dimensional conductance histogram which does not show distinctive peaks (Figure S1d).

It is well-established that the conductance of a nanocontact is mainly governed by the atoms at narrowest part of the constriction and therefore extremely sensitive to any minute variation in the chemical nature of the atoms there [11,12]. Due to the strong qualitative difference between the conductance histograms for gold-gold or carbon-gold, we rule out that the carbon tip is contaminated or wetted by gold atoms from the surface. We can therefore conclude that in the experiments in which gold is covered by molecules, the carbon tip apex is also not contaminated or wetted by gold atoms.

Additionally, the tip to sample coarse approach procedure of our home-build microscope can be configured to perform a very gentle tip engagement. The coarse approach is based on a piezoelectrically driven inertial motion and its speed is kept well below the speed of the tip-sample distance controller which is running during the coarse approach. In the reported experiments, we monitor the current during the coarse approach and we observe that it never exceeds 50 nA.

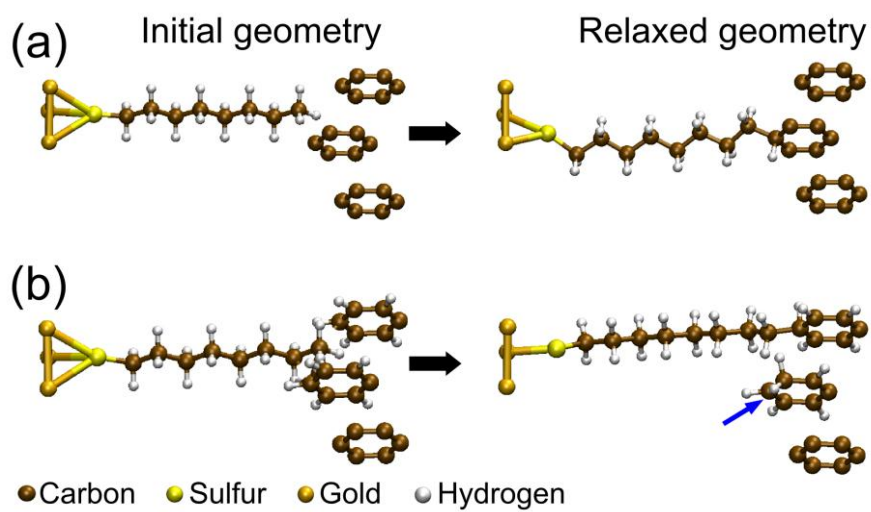

**Figure S2**. (a) The left structure shows the starting geometry considered in the DFT simulations, where a clean tip is simulated by three parallel carbon rings separated a distance of 3.345 Å, which corresponds to the equilibrium distance between bulk graphite planes. The molecule lies midway between the two upper carbon rings at a distance of the tip of 1.96 Å. The gold surface is simulated by a three-atom cluster and the molecule is bonded to it in a hollow position. The right structure shows the final geometry obtained after relaxing the structure. (b) The same as in panel (a), but for an initial geometry where the C atoms of the tip in the contact region have been saturated with H atoms. The last C atom of the alkane chain has penetrated inside the tip by a distance of 0.637 Å. The H atom transferred to the tip is indicated by a blue arrow.

**Density functional theory modelling**

The interaction between alkanes and extended carbon-based structures like graphite has been extensively studied [13,14]. However, the conclusions of those studies are not applicable to our low-dimensional situation. Therefore we have performed density functional theory (DFT) simulations to find a feasible mechanism for the formation of alkane-carbon contacts. Our discussion here is based on the analysis of a minimal model where a single molecule is coupled to a gold surface, which is modeled by a small cluster, and it interacts with a tip described by just three rings of six-membered C atoms that simulate the graphitic planes, see the left part of Figure S2(a). We have checked that increasing the size of the gold cluster or the tip does not change our main conclusions (see below). Within this simplified model, we have studied the interaction between the alkane chain and the carbon tip as a function of their separation. For this purpose, we first set up an initial geometry where the molecule is fully relaxed in front of the gold cluster and the tip is placed at a given distance from the molecule (in the examples of Figure S2 the molecule lies midway between two carbon rings). Then, we relax the structure keeping fixed the positions of both the tip and the Au atoms to obtain the equilibrium geometry. All the DFT calculations were done with the code TURBOMOLE v6.1. [15]. To be precise, we have used a split valence basis set with polarization functions and the BP86 exchange-correlation functional [16,17] in combination with the empirical approach of ref. [18] to describe the van der Waals interactions in our system.

Considering first that the tip is not chemically passivated, we find that for distances larger than about 1.97 Å there is a weak repulsion between the alkane chain and the carbon tip, i.e., no chemical bond is formed. However, if the tip is further approached, we find that the outermost C atom of the alkane chain can establish a single bond with one of the C atoms of the tip, see right structure of Figure S2(a). Notice that in this process, an H atom from the methyl group has been transferred to the tip. Our simulations suggest that if the molecule penetrates further inside the tip planes, a double bond could also be eventually established.

The discussion above suggests that even in the case of having initially a clean carbon tip, it could be passivated with hydrogen. This leads us to the question of whether it is possible to form a contact when the tip is decorated with hydrogen atoms, which is the most probable contaminant. In Figure S2(b) we show an example where in the starting geometry we have saturated all relevant dangling bonds in the carbon tip with H atoms. After relaxation, we find that, as in the case of a clean tip, the last C atom of the alkane chain can form a single bond by transferring an H atom to the tip. Notice that in this case the C atom of the tip involved in the bond changes its hybridization from $sp^2$ to $sp^3$. We find that in this case, the bond formation requires the molecule to partially penetrate inside the tip. On the other hand, let us mention that we have also studied the stretching process of these junctions by separating stepwise both the tip and the gold cluster. We find often that the junctions break at the molecule-gold interface, contrary to what is found in the experiments. This is a deficiency of our simple model that does not include the influence of the neighboring molecules in the SAM. It is known that the presence of other molecules in a SAM increases the binding energy per molecule, which makes more difficult to extract them from the substrate. This has been shown for the case of alkanethiols on gold in Ref [19].

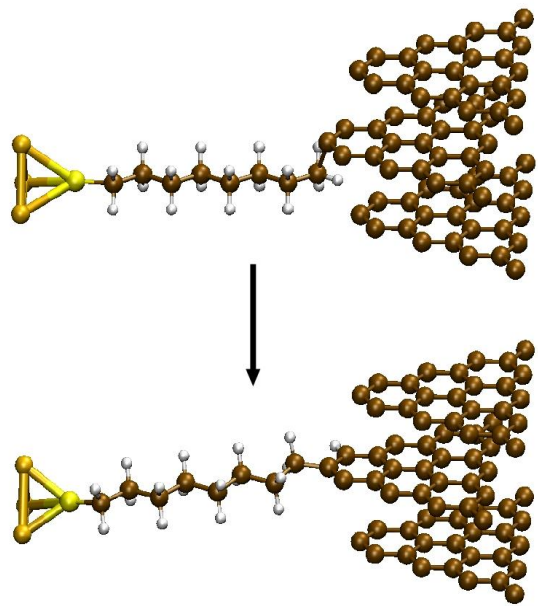

**Figure S3:** The upper structure shows the starting geometry considered in the DFT simulations, where a clean tip is simulated by three graphitic planes represented by six carbon rings. The planes are separated a distance equal to 3.354 Å and the molecule lies midway between the two lower planes at a distance of the tip of 1.795 Å. The gold surface is simulated by a three-atom cluster and the molecule is bonded to it in a hollow position. The lower structure shows the final geometry obtained after relaxing the structure. Notice the transfer of an H atom from the molecule to the tip.

We have also studied systematically whether the mechanism of the junction formation is sensitive to the size of the electrodes used in our DFT simulations. Thus for instance, we have carried out a series of simulations by increasing the size of the carbon electrode. One of these calculations is illustrated in Figure S3, where we have modeled the tip with up to six carbon rings per graphitic plane (to be compared with the single ring used in the simulations discussed above). As described before, we first set up an initial geometry where the molecule is fully relaxed in front of the gold cluster and the tip is placed at a given distance from the molecule (see upper picture in Figure S2). Then, we relax the structure keeping fixed the positions of both the tip and the Au atoms to obtain the equilibrium geometry. For the tip shown in Figure S3, we find that when the molecule-tip distance is less than approximately 1.8 Å, the outermost C atom of the alkane chain establishes a single bond with one of the C atoms of the tip by transferring an H atom to tip (see lower picture in Figure S3). This is exactly the same mechanism that we have found with the smaller tip (see above). This shows that the chemical bond originates from the local interactions between the outermost C atoms of both the molecule and the tip, whereas the actual shape of the tip does not play a crucial role. For this reason, and also for consistency, in the previous part we have stuck to the simplest geometry to discuss the different aspects, like the role of the tip passivation.

**References (for the supplementary information)**

[1] PAN based carbon fiber rope formed by 12000 individual fibers with a diameter of 7 μm. Manufactured by Hercules inc.: Reference: AS4-12K.

2. R. H. M. Smit, R. Grande, B. Lasanta, J. J. Riquelme, G. Rubio-Bollinger, and N. Agrait, Rev. Sci. Instrum. **78** (11), 113705 (2007).
3. DJ Johnson, J. Phys. D: Appl. Phys. **20**, 286 (1987).
4. A. Castellanos-Gomez, N. Agrait, and G. Rubio-Bollinger, Nanotechnology **21** (14), 145702 (2010).
5. G. E. Poirier and E. D. Pylant, Science **272** (5265), 1145 (1996).
6. G. Rubio, N. Agrait, and S. Vieira, Phys. Rev. Lett. **76** (13), 2302 (1996).
7. U Dürig, O Züger, LC Wang, and HJ Kreuzer, EPL (Europhysics Letters) **23**, 147 (1993).
8. C. Untiedt, M. J. Caturla, M. R. Calvo, J. J. Palacios, R. C. Segers, and J. M. Van Ruitenbeek, Phys. Rev. Lett. **98** (20), 206801 (2007).
9. C.A. Martin, D. Ding, J.K. Sørensen, T. Bjørnholm, J.M. van Ruitenbeek, and H.S.J. van der Zant, J. Am. Chem. Soc **130** (40), 13198 (2008).
10. C. Sirvent, JG Rodrigo, S. Vieira, L. Jurczyszyn, N. Mingo, and F. Flores, Phys. Rev. B **53** (23), 16086 (1996).
11. G. Rubio-Bollinger, C. de Las Heras, E. Bascones, N. Agrait, F. Guinea, and S. Vieira, Phys. Rev. B **67** (12), 121407 (2003).
12. E. Scheer, N. Agraït, J.C. Cuevas, A.L. Yeyati, B. Ludoph, A. Martín-Rodero, G.R. Bollinger, J.M. van Ruitenbeek, and C. Urbina, Nature **394** (6689), 154 (1998).
13. T. Yang, S. Berber, J. F. Liu, G. P. Miller, and D. Tománek, J. Chem. Phys. **128**, 124709 (2008).
14. B. Ilan, G. M. Florio, M. S. Hybertsen, B. J. Berne, and G. W. Flynn, Nano Lett. **8** (10), 3160 (2008).
15. R. Ahlrichs, M. Bär, M. Häser, H. Horn, and C. Kölmel, Chem. Phys. Lett. **162** (3), 165 (1989).
16. J. P. Perdew, Phys. Rev. B **33** (12), 8822 (1986).
17. A. D. Becke, Phys. Rev. A **38** (6), 3098 (1988).
18. S. Grimme, J. Comput. Chem. **25** (12), 1463 (2004).
19. S.N. Patole, C.J. Baddeley, D. O'Hagan, N.V. Richardson, F. Zerbetto, L.A. Zotti, G. Teobaldi, and W.A. Hofer, J. Chem. Phys. **127**, 024702 (2007).